\documentclass[applemac]{aa}
\usepackage{natbib}
\usepackage{txfonts}
\usepackage{graphicx}
\usepackage{color}
\bibpunct{(}{)}{;}{a}{}{,}

\def\ammo{\rm NH_3}
\def\dammo{\rm NH_2D}
\def\ddammo{\rm NHD_2}
\def\ndthree{\rm ND_3}

\def\diaz{\rm N_2H^+}

\def\htwo{\rm H_2}

\def\hthree{\rm H_3^+}
\def\htwod{\rm H_2D^+}
\def\dtwoh{\rm D_2H^+}
\def\dthree{\rm D_3^+}

\def\ammohplus{\rm NH_4^+}
\def\dammohplus{\rm NH_3D^+}
\def\ddammohplus{\rm NH_2D_2^+}
\def\ndthreehplus{\rm NHD_3^+}

\def\ammoplus{\rm NH_3^+}
\def\dammoplus{\rm NH_2D^+}
\def\ddammoplus{\rm NHD_2^+}
\def\ndthreeplus{\rm ND_3^+}

\def\htwo{\rm H_2}
\def\el{\rm e^-}
\def\hplus{\rm H^+}
\def\arrow{\mathop{\rightarrow}\limits}

\begin{document}

\title{Spin-state chemistry of deuterated ammonia}
\author{O. Sipil\"a\inst{1},
		J. Harju\inst{1,2},
		P. Caselli\inst{1},
	     \and{S. Schlemmer\inst{3}}
}
\institute{Max-Planck-Institute for Extraterrestrial Physics (MPE), Giessenbachstr. 1, 85748 Garching, Germany \\
e-mail: \texttt{osipila@mpe.mpg.de}
\and{Department of Physics, P.O. Box 64, 00014 University of Helsinki, Finland}
\and{I. Physikalisches Institut, Universit\"at zu K\"oln, Z\"ulpicher Strasse 77, 50937 K\"oln, Germany}
}

\date{Received / Accepted}

\abstract{}
{We aim to develop a chemical model that contains a consistent description of spin-state chemistry in reactions involving chemical species with multiple deuterons. We apply the model to the specific case of deuterated ammonia, to derive values for the various spin-state ratios.}
{We apply symmetry rules in the complete scrambling assumption to calculate branching ratio tables for reactions between chemical species that include multiple protons and/or deuterons. New reaction sets for both gas-phase and grain-surface chemistry are generated using an automated routine that forms all possible spin-state variants of any given reaction with up to six H/D atoms, using the pre-determined branching ratios. Single-point as well as a modified Bonnor-Ebert model are considered to study the density and temperature dependence of ammonia and its isotopologs, and the associated spin-state ratios.}
{We find that the spin-state ratios of the ammonia isotopologs are, at late times, very different from their statistical values. The ratios are rather insensitive to variations in the density, but present strong temperature dependence. We derive high peak values ($\sim 0.1$) for the deuterium fraction in ammonia, in agreement with previous (gas-phase) models. The deuterium fractionation is strongest at high density, corresponding to a high degree of depletion, and also presents temperature dependence. We find that in the temperature range 5 to 20\,K, the deuterium fractionation peaks at $\sim 15$\,K while most of the ortho/para (and meta/para for $\rm ND_3$) ratios present a minimum at 10\,K (ortho/para $\rm NH_2D$ has instead a maximum at this temperature).}
{Owing to the density and temperature dependence found in the abundances and spin-state ratios of ammonia and its isotopologs, it is evident that observations of ammonia and its deuterated forms can provide important constraints on the physical structure of molecular clouds.}


\maketitle

\section{Introduction}

The chemical evolution of cold molecular cloud cores is characterized by increasing abundances of the nitrogen-containing species $\diaz$ and $\ammo$, and by the enhancement of deuterium fractionation (e.g., \citealt{2000A&A...356.1039T}; \citealt{2002ApJ...569..815T}). The three isotopologs of deuterated ammonia, $\dammo$, $\ddammo$, and $\ndthree$, look therefore particularly useful indicators of temporal changes in the chemical composition of starless and prestellar
cores. Besides that the $\dammo/\ammo$, $\ddammo/\dammo$, $\ndthree/\ddammo$ abundance ratios are likely to increase with time (\citealt{2006A&A...456..215F}; \citealt{Roueff15}), also the ratios of different nuclear spin isomers of these molecules can vary. The reactions leading to ammonia and its deuterated forms are recapitulated in \cite{2001ApJ...553..613R}. Deuteration occurs primarily in reactions between $\ammo$ and deuterated ions. In dense, cold clouds this means mainly reactions with the deuterated forms of $\hthree$.  Through these reactions, the nuclear spin ratios of $\dammo$, $\ddammo$, and $\ndthree$ become dependent on the nuclear spin ratios of $\htwod$, $\dtwoh$, and $\dthree$. The latter ratios depend on the $\htwo$ {\sl ortho}:{\sl para} ratio, hereafter o/p-$\htwo$ (e.g., \citealt{WFP04}), which also influences the formation rate of ammonia \citep{LeBourlot91, Wirstrom12}. The o/p-$\htwo$ ratio is predicted to decrease drastically as the core evolves (e.g., \citealt{1984MNRAS.209...25F}; \citealt{2011ApJ...739L..35P}; \citealt{2014Natur.516..219B}).
While the formation rates of different spin isomers of ammonia probably vary because of competing ion-molecule reactions, grain-surface reactions should produce them in ratios of the nuclear spin statistical weights because H and D atoms are added one by one. However, the accretion of ammonia molecules synthesized in the gas phase leads to gradual changes 
of the average spin ratios on grains.

The ground-state rotational transitions of the {\sl ortho} and {\sl para} species of $\dammo$ and $\ddammo$, as well as those of the {\sl meta} and {\sl para} species of $\ndthree$ are observable with ground-based telescopes (\citealt{2001ApJ...554..933S}; \citealt{2002ApJ...571L..55L};  \citealt{2002A&A...388L..53V}; \citealt{Roueff05}; \citealt{2006A&A...454L..63G};  \citealt{2006ApJ...636..916L}).\footnote{The fundamental rotation-inversion lines of {\sl ortho}-$\ndthree$ at 615 and 618 GHz may be observable from the ground in extremely good conditions.} In the few sources where these have been determined, the $\dammo/\ammo$ and $\ddammo/\dammo$ column density ratios have been found to lie in the range $\sim 0.1-0.2$, whereas the $\ndthree/\ddammo$ is somewhat lower, $\sim 0.02 - 0.1$, implying that $\ndthree/\ammo \sim 10^{-4}-10^{-3}$ (\citealt{Roueff05}; \citealt{2006A&A...454L..63G}; \citealt{2006ApJ...636..916L}). These numbers signify enormous deuterium enhancements in ammonia with respect to the cosmic D/H abundance ratio. The observed spin ratios are consistent with the nuclear spin statistical weights, taking observational uncertainties into account. With the increasing sensitivity of (sub)millimetre telescopes, it will be possible to improve the accuracy of these measurements, and extend the investigation to cover larger and more varied samples of dense cores.

A valid description of chemistry in dense cores should be able to explain, for example, the observed abundances of the different deuterated isotopologs and spin isomers of ammonia, which belongs to the most useful probes of dense molecular cloud cores.  With this goal in mind, we have developed a comprehensive gas-grain chemistry model which includes deuterium and the spin-state chemistry of molecules containing multiple H and/or D nuclei. In an earlier version of this model, the deuterium fractionation and spin selection rules were applied to light molecules \citep{Sipila13}. The model has recently been extended to include heavier deuterated molecules and to take into account the spin selection rules for hydrogen in the formation of ammonia and water (\citealt{Sipila15}, hereafter S15). In the present update, we implement the spin selection rules for reactions involving up to six hydrogen and deuterium nuclei.

As described in Sect.\,2 of this paper, the inclusion of the spin-state chemistry of multiply deuterated species has required a complete revision of the spin separation routine which is now based on the symmetries of the nuclear spin functions rather than angular momentum algebra used in the previous versions of the model. In Sect.\,3, we predict the abundances of ammonia and its deuterated forms, and the spin ratios of these molecules, as functions of time in homogenous models corresponding to dark cloud and dense core conditions. In Sect.\,4, we discuss the temperature dependence of the deuteration of ammonia, and derive radial abundance profiles in two modified Bonnor-Ebert sphere models. In this Section, we also compare our predictions with the gas phase chemical model of \citet{Roueff05}. Finally, in Sect.\,5 we present our conclusions. An online appendix of this paper contains a couple of examples of how to calculate the nuclear spin branching ratios using symmetry rules.

\section{Model}

\subsection{Theory}

In the present model, the different nuclear spin symmetry states of molecules containing two or more hydrogen and/or deuterium nuclei are treated as separate species. This implies that for example for the doubly deuterated ammonium ion, $\rm NH_2D_2^+$, we keep track of four different forms depending on the two possible spin symmetries, ``ortho'' ($A$) and ``para'' ($B$), of the pairs of the H and D nuclei. In reactions involving multiple H or D nuclei, the branching ratios between different nuclear spin channels are assumed to be determined by their pure nuclear spin statistical weights, i.e., the number of nuclear spin functions belonging to each channel, which are calculated using permutation symmetry algebra (\citealt{Quack77}; \citealt{Park07}; \citealt{Hugo08}; \citealt{Hugo09}). The method uses the assumptions of nuclear spin conservation and full scrambling of nuclei in reactive collisions. The branching ratios are calculated for reactions involving up to six hydrogen or deuterium nuclei.

In our previous papers (\citealt{Sipila13}; S15), the nuclear spin branching ratios were deduced using the method of \cite{Oka04} based on angular momentum algebra. The same method was applied to nuclear-spin selection rules of nitrogen hydrides by \citet{Rist13}. For reactions where several deuterium nuclei are involved, Oka's method is, however, impractical, because species are distinguished by their nuclear spin symmetries, but there is no one-to-one correspondence between angular momentum and symmetry representations for systems with three or more D nuclei (\citealt{Hugo08}). A comparison between the total angular momentum and the total symmetry representations of systems containing up to five spin 1/2 and spin 1 particles is presented in Table~2.13 of \cite{Hugo08}. For multiple H nuclei (or any system of spin 1/2 particles), the method using symmetry rules gives the same results as Oka's method, because for these systems each value of the spin angular momentum is associated with one and only one symmetry representation. Also in the case of multiple D nuclei, the total nuclear spin is conserved when using the symmetry rules. This could be seen from branching ratio tables if the symmetry species were subdivided according to the total nuclear spin (an example of this is given in \citealt{Hugo08}; Table 2.1). 

For the calculation of the statistical branching ratios using symmetry rules, one needs to determine the possible symmetry species (i.e., representations) of the reactants ($\Gamma_i\otimes\Gamma_j$; where $\otimes$ denotes a direct product), the intermediate complex ($\Gamma_n$), and the products ($\Gamma_k\otimes\Gamma_l$, $i+j=n=k+l$), in appropriate permutation groups ($S_i\otimes S_j$, $S_n$, and $S_k\otimes S_l$). In addition, one needs to derive the correlation tables between the group of the complex and its subgroups representing the reactants and products. This is done applying the standard methods of group theory. The correlation tables tell us which nuclear spin symmetry species of the reaction complex are feasible given the symmetries of the reactants, and which symmetry states of the products are possible when the complex decays (cf. \citealt{Quack77}; Eqs.\,(12)-(15)). In this description, the statistical weight of the product $\Gamma_k\otimes\Gamma_l$ is given by $f(\Gamma_k\otimes\Gamma_l)\times {\rm dim}(\Gamma_k)\times{\rm dim}(\Gamma_k)$, where $f(\Gamma_k\otimes\Gamma_l)$ is the frequency of $\Gamma_k\otimes\Gamma_l$ in the total symmetry representation generated by the nuclear spin functions of the product molecules in $S_k\otimes S_l$, and ${\rm dim}(\Gamma)$ is the dimension of the representation $\Gamma$. Besides being proportional to the number of nuclear spin functions 
belonging to each channel, branching ratios calculated in this manner are proportional to the number of rovibronic states occupied by these species in the high temperature limit (\citealt{2007JChPh.126d4305P}, their Appendix~A). A couple of examples of the procedure are given in the Appendix. The determination of the symmetries of the nuclear spin wave functions is explained lucidly in \cite{Bunker79} and in \cite{BunkerJensen06}. In this work, where we do not consider the rovibronic states of the molecules, it is sufficient to classify the nuclear spin functions in the complete nuclear spin permutation group which does not include the inversion operation, see also \citet{Crabtree13}. Most of the statistical branching ratio tables needed in the present study are available in Tables~III and IV of \citet{Hugo09}. In order to cover also reactions with six identical nuclei, like $\rm NH_3 + H_3^+$ and $\rm ND_3 + D_3^+$, and to transfer the spin selection rules to the chemical network, we developed an automated routine for producing these tables. The character tables for permutation groups needed in this process were obtained using \cite{GAP4}\footnote{GAP is a system for computational group theory, http://www.gap-system.org}.

\subsection{Application}

We designed a new routine for the spin-state separation of chemical reactions involving multiple protons and/or deuterons, utilizing the group-theoretical approach described above. As opposed to the model described in S15, the new routine produces consistent branching ratios for reactions involving deuterated species, instead of copying the spin-state separation rules from the pure-hydrogen counterpart reactions.

Because hydrogen and deuterium are distinguishable, we can divide each reaction involving both hydrogen and deuterium into two parts. For example, the reaction
\begin{equation}\label{eq:example}
\rm CH_2D_2^+ + HDO \longrightarrow \left( COH_3D_3^+ \right)^{\ast} \longrightarrow H_2DO^+ + CHD_2
\end{equation}
is divided into
\begin{eqnarray}
\rm H_2 + H &\longrightarrow& \rm \left( H_3 \right)^{\ast} \longrightarrow \rm H_2 + H \label{eq:h_ex} \\ 
\rm D_2 + D &\longrightarrow& \rm \left( D_3 \right)^{\ast} \longrightarrow \rm D + D_2 \label{eq:d_ex} \, ,
\end{eqnarray}
where we have ignored the carbon and oxygen atoms, and the charge. The selection rules for hydrogen split reaction (\ref{eq:h_ex}) into the following branches:
\begin{eqnarray*}
\rm oH_2 + H &\mathop{\longrightarrow}\limits^{\frac{5}{6}}& {\rm oH_2} + {\rm H} \\
 &\mathop{\longrightarrow}\limits^{\frac{1}{6}}& {\rm pH_2} + {\rm H} \\
\rm pH_2 + H &\mathop{\longrightarrow}\limits^{\frac{1}{2}}& {\rm oH_2} + {\rm H} \\
 &\mathop{\longrightarrow}\limits^{\frac{1}{2}}& {\rm pH_2} + {\rm H} \, ,
\end{eqnarray*}
while the selection rules for deuterium split reaction (\ref{eq:d_ex}) into
\begin{eqnarray*}
\rm oD_2 + D &\mathop{\longrightarrow}\limits^{\frac{7}{9}}& {\rm D} + {\rm oD_2} \\
 &\mathop{\longrightarrow}\limits^{\frac{2}{9}}& {\rm D} + {\rm pD_2} \\
\rm pD_2 + D &\mathop{\longrightarrow}\limits^{\frac{4}{9}}& {\rm D} + {\rm oD_2} \\
 &\mathop{\longrightarrow}\limits^{\frac{5}{9}}& {\rm D} + {\rm pD_2} \, .
\end{eqnarray*}
We then combine the hydrogen and deuterium branches by multiplying all possible combinations together, and we end up with $4 \times 4 = 16$ different branches for reaction~(\ref{eq:example}). Note that as a result, $\rm CH_2D_2^+$ receives a composite spin made up of hydrogen and deuterium spin states, for example, $\rm CoH_2pD_2^+$.

The spin-state separation routine presented here allows us to explicitly consider the spin states of all species which contain multiple H and/or D atoms. However, we do not track the spin states of all species because this would lead to an enormous reaction network, with a total reaction count far into the hundreds of thousands. Instead, we adopt the averaging process described in S15, in which we assume a 1:1 spin-state abundance ratio for those species whose spin states we do not want to keep track of. We have verified that this assumption has negligible influence on the results presented below. In this paper, we focus on the spin-state chemistry of ammonia and its deuterated forms. We also keep the spin states of the species included in the formation network of water, so that we can compare our new model to that presented in S15. In reaction~(\ref{eq:example}), the only species for which we retain the spin states is $\rm H_2DO^+$, and the averaging process finally results in two branches:
\begin{eqnarray*}
\rm CH_2D_2^+ + HDO &\mathop{\longrightarrow}\limits^{\frac{2}{3}}& {\rm oH_2DO^+} + {\rm CHD_2} \\
 &\mathop{\longrightarrow}\limits^{\frac{1}{3}}& {\rm pH_2DO^+} + {\rm CHD_2} \, .
\end{eqnarray*}

The separation process is fully automated, so any reaction involving multiple protons in our base reaction set (see S15) is automatically deuterated and spin-state separated according to the selection rules discussed above. The only exception to this is charge transfer reactions, for which we assume that the spin states are conserved in the reaction. The reaction $\rm N^+ + H_2$ initiating the ammonia synthesis has been treated in the same manner as done in \citet{Dislaire12} and \citet{Roueff15}, using the experimental data from \citet{Marquette88}. In addition to the updates listed in S15, the 
new reaction set includes the rate coefficients for $\rm N^+ + HD$ and $\rm N^+ + D_2$ from Marquette et al., taking into account the slightly different ground-state energies of o$\rm D_2$ and p$\rm D_2$ in the reaction enthalpies. For the reactions $\rm NH_3^+ + H_2 \rightarrow NH_4^+ + H$, we adopt the rate coefficients from \citet{LeGal14}, which correspond to those used in \citet{Roueff05}.

In what follows, we show results calculated with the gas-grain chemical code discussed in S15. In summary, we choose the {\tt osu\_01\_2009} reaction set as the basis set which is deuterated and spin-state separated according to the approaches discussed above. The initial grain-surface chemistry reaction set is adopted from \citet{Semenov10}, and is similarly deuterated and spin-state separated. A complete description of the model including adopted model parameters and updates to the (basis) reaction sets is presented in S15.

\section{Results}

\subsection{Comparison to S15}

\begin{figure}
\centering
\includegraphics[width=\columnwidth]{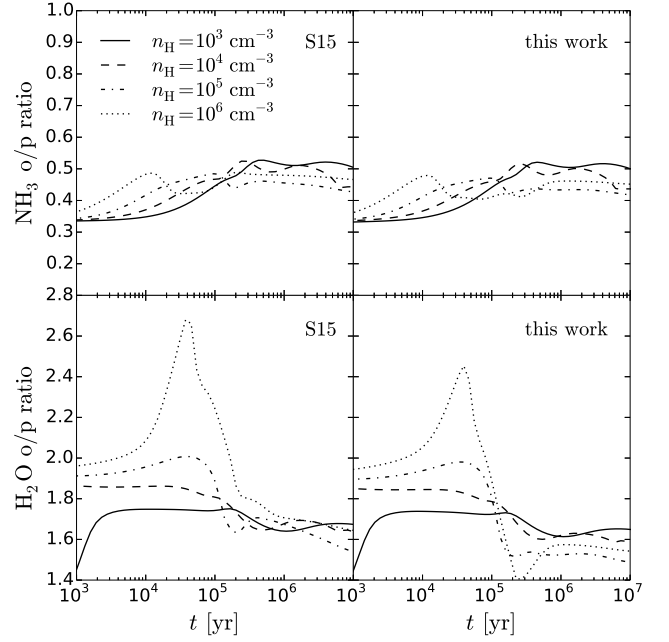}
\caption{Spin-state ratios of $\rm NH_3$ ({\sl upper panels}) and $\rm H_2O$ ({\sl lower panels}) as functions of time at different densities, labeled in the upper left panel. The left-hand panels present results calculated with the (gas-grain) model of S15, while the right-hand panels correspond to the present work.
}
\label{fig:s15test}
\end{figure}

Because of the isomorphism between the symmetry and angular momentum representations for hydrogenated species, we recover for these the same branching ratios in the present model as with the method of \citet{Oka04} employed in S15. However, the branching ratios in reactions involving deuterated species are in many cases significantly different between the two models.

In Fig.\,\ref{fig:s15test}, we compare the o/p ratios of ammonia and water predicted by the S15 model and the present model. The temperature is set to $T = 10$\,K, and the other physical parameters are the same as in S15. The addition of spin states for multiply deuterated species has in general little effect on the spin-state ratios of undeuterated species, but some difference is evident, especially for water at high density. Inspection of the reaction rates reveals that the differences between the models seen in both the ammonia and water o/p ratios are mainly caused by the electron recombinations of (o/p) $\rm NH_3D^+$ and (o/p) $\rm H_2DO^+$. The S15 model contains some pathways for these reactions that are forbidden by the proper selection rules adopted in this work. Nevertheless, the difference between the models is small, and we have checked that this agreement holds for other (undeuterated) species as well.

The results shown in the following sections were calculated with the S15 chemical code, adopting the new reaction sets developed for this paper. Like in the test above, the temperature is in all calculations set to $T = 10$\,K, and the other physical parameters are the same as in S15.

\subsection{Deuterium fractionation of ammonia} \label{ss:ammoniadfrac}

\begin{figure*}[hbt]
\centering
   \unitlength=1.0mm
  \begin{picture}(150,50)(0,0)

\color{black} 
\put(10,10){\circle{11}}
\put(10,10){\makebox(0,0){$\ammoplus$}}

\put(20,40){\circle{11}}
\put(20,40){\makebox(0,0){$\ammohplus$}}

\color{blue} 
\put(30,10){\circle{11}}
\put(30,10){\makebox(0,0){$\ammo$}}

\color{black} 
\put(50,10){\circle{11}}
\put(50,10){\makebox(0,0){$\dammoplus$}}

\color{black} 
\put(60,40){\circle{11}}
\put(60,40){\makebox(0,0){$\dammohplus$}}

\color{blue} 
\put(70,10){\circle{11}}
\put(70,10){\makebox(0,0){$\dammo$}}

\color{black} 
\put(90,10){\circle{11}}
\put(90,10){\makebox(0,0){$\ddammoplus$}}

\color{black} 
\put(100,40){\circle{11}}
\put(100,40){\makebox(0,0){$\ddammohplus$}}

\color{blue} 
\put(110,10){\circle{11}}
\put(110,10){\makebox(0,0){$\ddammo$}}

\color{black} 
\put(130,10){\circle{11}}
\put(130,10){\makebox(0,0){$\ndthreeplus$}}

\put(140,40){\circle{11}}
\put(140,40){\makebox(0,0){$\ndthreehplus$}}

\color{blue} 
\put(150,10){\circle{11}}
\put(150,10){\makebox(0,0){$\ndthree$}}

\color{black}
\thinlines
\put(23,10){\vector(-1,0){6}}
\put(63,10){\vector(-1,0){6}}
\put(103,10){\vector(-1,0){6}}
\put(143,10){\vector(-1,0){6}}

\put(10,17){\vector(1,3){6}}
\put(50,17){\vector(1,3){6}}
\put(90,17){\vector(1,3){6}}
\put(130,17){\vector(1,3){6}}

\put(50,17){\vector(-4,3){25}}
\put(90,17){\vector(-4,3){25}}
\put(130,17){\vector(-4,3){25}}

\put(23,34){\vector(1,-3){6}}
\put(63,34){\vector(1,-3){6}}
\put(103,34){\vector(1,-3){6}}
\put(143,34){\vector(1,-3){6}}

\thinlines
\color{blue}
\put(31,17){\vector(-1,3){6}}
\put(71,17){\vector(-1,3){6}}
\put(111,17){\vector(-1,3){6}}
\put(151,17){\vector(-1,3){6}}

\color{blue}
\put(31,17){\vector(1,1){20}}
\put(31,17){\vector(3,1){57}}
\put(31,17){\line(5,1){100}}

\put(71,17){\vector(1,1){20}}
\put(71,17){\vector(3,1){60}}

\put(111,17){\vector(1,1){20}}

\thinlines
\color{black}
\put(24,22){\makebox(0,0){$\el$}}
\put(64,22){\makebox(0,0){$\el$}}
\put(104,22){\makebox(0,0){$\el$}}
\put(144,22){\makebox(0,0){$\el$}}
\color{black}
\put(14,18){\makebox(0,0){$\htwo$}}
\put(35,32){\makebox(0,0){$\htwo$}}
\put(54,18){\makebox(0,0){$\htwo$}}
\put(94,18){\makebox(0,0){$\htwo$}}
\put(134,18){\makebox(0,0){$\htwo$}}

\put(20,6){\makebox(0,0){$\hplus$}}
\put(60,6){\makebox(0,0){$\hplus$}}
\put(100,6){\makebox(0,0){$\hplus$}}
\put(140,6){\makebox(0,0){$\hplus$}}

\color{blue}
\put(31,27){\makebox(0,0){$\hthree$}}
\put(44,38){\makebox(0,0){$\htwod$}}
\put(77,38){\makebox(0,0){$\dtwoh$}}
\put(117,38){\makebox(0,0){$\dtwoh$, $\dthree$}}

\color{black}
\end{picture}
\caption{Principal reactions determining the abundances and spin ratios of deuterated ammonia isotopologs at late stages of chemical evolution in cold, dense cores. The reactions ${\rm NH_{3-X} D_X + H_{3-Y} D_Y^+}$ are indicated with blue arrows.}
\label{fig:damm_cycle}
\end{figure*}
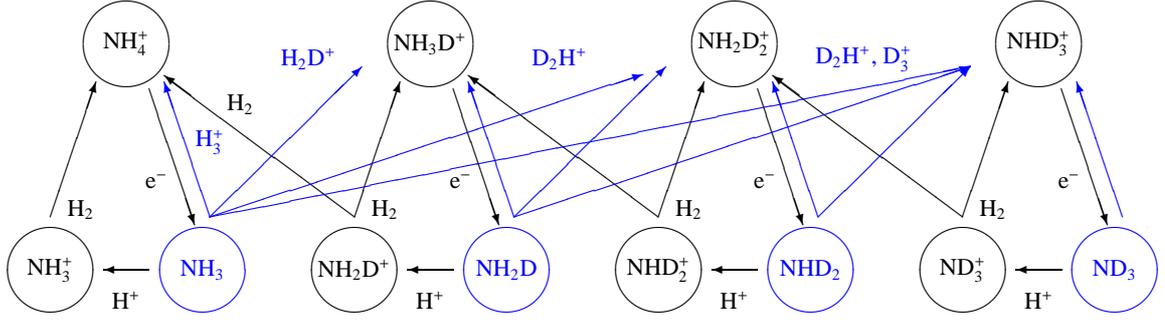

By inspecting the production and destruction rates of deuterated forms of ammonia at various stages of the simulation, one finds that at late times the deuteration of ammonia proceeds as outlined in the steady-state model of \cite{2001ApJ...553..613R}. The $\ammo$ abundance has already built up when the gas-phase deuteration becomes effective, and deuteron transfer reactions between ammonia and deuterated ions form, e.g., $\dammohplus$ which can recombine with an electron to give $\dammo$, etc. In the present model, however, $\htwod$, $\dtwoh$, and $\dthree$ are the most important sources of deuterons at late times. The dominant reactions forming $\dammo$, $\ddammo$, and $\ndthree$ are shown in Fig.~\ref{fig:damm_cycle}. This diagram, which is an extension to the schematic $\ammo$ cycle shown in Fig.~A.1. of S15, should be compared with Fig.~1 of \cite{2001ApJ...553..613R}. Besides the deuteration pathways, the diagram shows the constant recycling between $\ammo$, $\ammoplus$, and $\ammohplus$, and similar cyclic processes among their deuterated variants, caused by charge transfer reactions with $\hplus$. This recycling tends to reduce the deuterium fractionation, because $\dammoplus + \htwo$, for example, leads also to $\ammohplus$ (we assume a statistical branching ratios, in this case 4:1 in favour of $\dammohplus$). The reaction ${\rm N^+} + {\rm HD} \rightarrow {\rm ND^+} + {\rm H}$ with a very small endothermicity \citep{Marquette88} speeds up the deuteration of ammonia at early stages, but this pathway has no influence on the abundances at the deuterium peak.
 
Figure\,\ref{fig:ammonia_abus} shows the total abundances, i.e., sums over spin states, of ammonia and its deuterated forms. At late time at high density, ammonia starts to deplete from the gas phase (see also S15). For the various forms of deuterated ammonia, an abundance peak is observed which coincides with the initiation of deuterium chemistry owing to CO depletion. Like ammonia, its isotopologs also deplete from the gas phase, more strongly than ammonia itself. This is because of the depletion of deuterated molecules (including the ammonia isotopologs) onto grain surfaces which traps deuterium onto the surfaces; this results ultimately in HD depletion and consequently the overall gas-phase deuteration efficiency decreases (\citealt{Sipila13}; S15).

The abundance ratios of ammonia and its isotopologs are presented in Fig.\,{\ref{fig:ammonia_abu_ratios}. The deuterium fractionation is highest at high density where depletion is strongest. The relative depletion of the ammonia isotolopogs is rather uniform, with all abundance ratios ultimately dropping by a factor of 3-4 from their peak values. A high degree of fractionation is observed, and the peak values for all of the ratios are $\gtrsim 0.1$. This issue is further discussed in Sect.\,\ref{ss:roueffcomparison}, where we compare our gas-grain modeling results to the gas-phase models of \citet{Roueff05}.

\subsection{The spin-state ratios of deuterated ammonia}\label{ss:spinratios}

\begin{table}
\caption{Spin symmetries of molecules with 2-4 H or D nuclei. The numbers in front of the symmetry labels indicate their frequencies in the total nuclear spin representation of the complex. The values of the possible nuclear spin angular momenta, $I$, and the nuclear spin statistical weights, $g_I$, are given below each symmetry species, as well as their "para", "ortho", and "meta" 
appellations, when applicable.}
\begin{minipage}{4.0cm}
\begin{tabular}{c|cc}
${\rm H_2}$  &  $3A$   &$B$   \\ \hline
              &  $I=1$  & $I=0$      \\
              &  $g_I=3$& $g_I=1$  \\
              &  {\sl ortho}  & {\sl para}  \\
\end{tabular} \\
\end{minipage}
\begin{minipage}{3.8cm}
\begin{tabular}{c|cc}
${\rm D_2}$ &  $6A$   & $3B$ \\ \hline
               &  $I=0,2$ & $I=1$ \\
                & $g_I=6$  & $g_I=3$ \\
                & {\sl ortho}    & {\sl para} \\
\end{tabular}
\end{minipage}
\newline
\newline
\begin{minipage}{4.0cm}
\begin{tabular}{c|cc}
${\rm H_3}$ & $4A_1$   & $2E$   \\ \hline
             & $I=3/2$ & $I=1/2$ \\
             & $g_I=4$ & $g_I=4$  \\
             & {\sl ortho}   & {\sl para}  \\
\end{tabular} \\
\end{minipage}
\begin{minipage}{3.8cm}
\begin{tabular}{c|ccc}
${\rm D_3}$ & $10A_1$&$A_2$&$8E$ \\ \hline
        &$I=1,3$& $I=0$  &$I=1,2$ \\
            &$g_I=10$& $g_I=1$ &$g_I=16$\\
             & {\sl meta} &  {\sl para} & {\sl ortho} \\
\end{tabular}
\end{minipage}
\newline
\newline
\begin{tabular}{c|ccc}
${\rm H_4}$ & $5A_1$&$E$&$3F_1$ \\ \hline
             & $I=2$ & $I=0$ & $I=1$ \\
             & $g_I=5$& $g_I=2$ & $g_I=9$ \\
             & {\sl meta}  & {\sl para} & {\sl ortho} \\
\end{tabular}
\newline
\newline
\begin{tabular}{c|cccc}
  ${\rm D_4}$ & $15A_1$ & $6E$ & $15F_1$ & $3F_2$\\ \hline
               & $I=0,2,4$ & $I=0,2$ & $I=1,2,3$ & $I=1$ \\
               & $g_I=15$ & $g_I=12$ & $g_I=45$ & $g_I=9$ \\
\end{tabular}
       \label{table:spin_species}
\end{table}

In Fig.\,\ref{fig:ammonia_ratios}, we show the spin-state ratios of the deuterated forms of ammonia. All of the ratios show dependence on both time and density. For all of the ammonia isotopologs, the most important formation mechanism is the electron recombination of the (singly or multiply) deuterated ammonium ion. To complement Fig.\,\ref{fig:ammonia_ratios}, we show in
Fig.\,\ref{fig:ammonia_ions} the spin-state ratios of selected ions. In the following, we quote the spin symmetries of the various species; these are tabulated in Table \ref{table:spin_species}.

\begin{figure}
\centering
\includegraphics[width=\columnwidth]{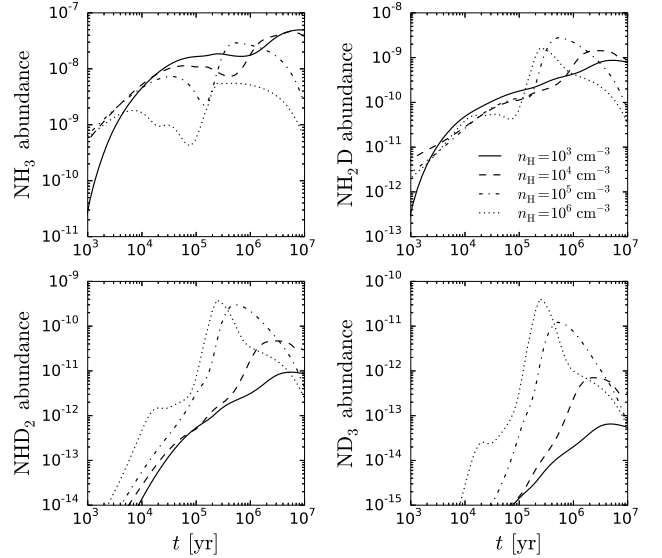}
\caption{Abundances of ammonia and its deuterated forms as functions of time at different densities, labeled in the upper right panel.
}
\label{fig:ammonia_abus}
\end{figure}

\begin{figure*}
\centering
\includegraphics[width=2.0\columnwidth]{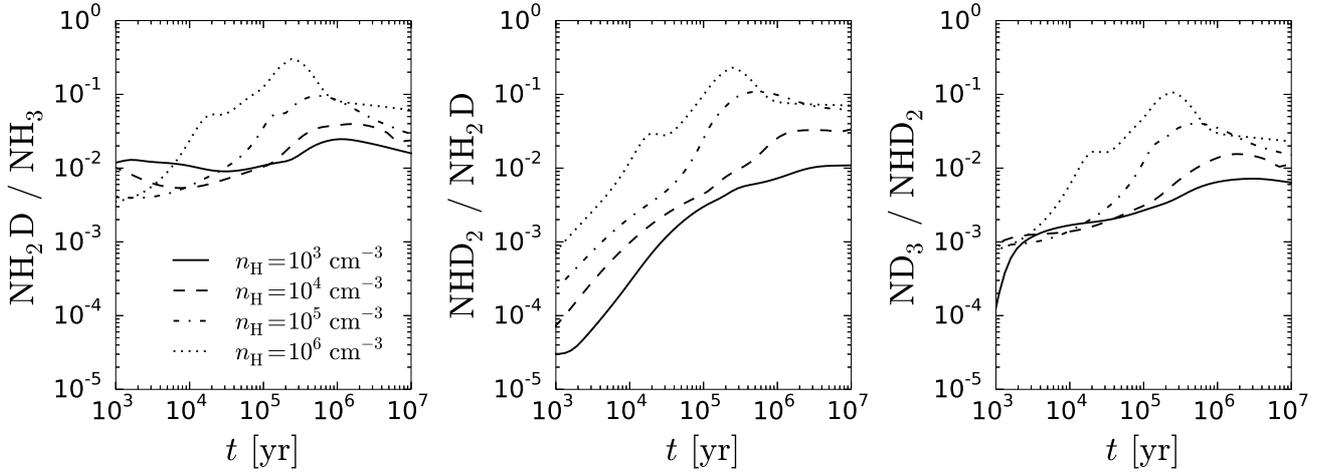}
\caption{Abundance ratios of ammonia and its deuterated forms as functions of time at different densities, labeled in the left-hand panel.
}
\label{fig:ammonia_abu_ratios}
\end{figure*}

\begin{figure}
\centering
\includegraphics[width=\columnwidth]{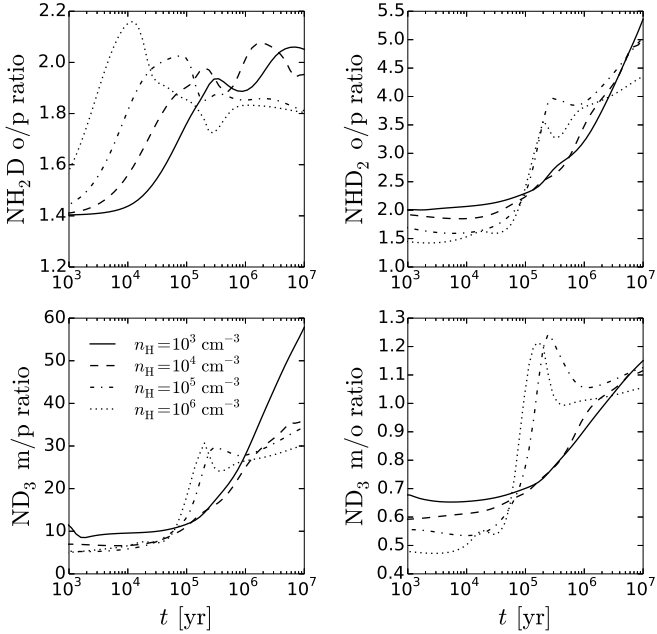}
\caption{Spin-state ratios of the deuterated forms of ammonia as functions of time at different densities, labeled in the lower left panel.}
\label{fig:ammonia_ratios}
\end{figure}

At late times, the ortho:para ratio of $\dammo$ is largely determined by the cycle $\dammo \arrow^{\hplus} \dammoplus \arrow^{\htwo} \dammohplus \arrow^{\el} \dammo$ (see Fig.~\ref{fig:damm_cycle}). The spin ratio of $\ammo$ was discussed in Appendix~A of S15. The electron recombination of $\rm NH_3D^+$ yields ortho and para $\rm NH_2D$ with the following branching ratios:
\begin{eqnarray*}
\rm oNH_3D^+ + e^- &\mathop{\longrightarrow}\limits^{k_1}& 
                      {\rm oNH_2D} + {\rm H} \\
\rm pNH_3D^+ + e^- &\mathop{\longrightarrow}\limits^{\frac{1}{2}k_1}& 
                     {\rm pNH_2D} + {\rm H} \\
                  &\mathop{\longrightarrow}\limits^{\frac{1}{2}k_1}& 
                     {\rm oNH_2D} + {\rm H} \, .
\end{eqnarray*}
Since the destruction rates of o$\dammo$ and p$\dammo$ (mainly through reactions with $\hplus$ and $\hthree$) are similar, we get o/p-$\dammo \sim 1 + 2$o/p-$\dammohplus$. Moreover, because the o/p ratio of $\rm NH_3D^+$ is $\sim~0.4-0.5$ at late times depending on the density (Fig.\,\ref{fig:ammonia_ions}), o/p$\rm NH_2D$ should obtain the values 2 (low density) or 1.8 (high density), in agreement with Fig.\,\ref{fig:ammonia_ratios}.  

The o/p ratios of $\dammo$ and $\dammohplus$ are, however, closely coupled because many of the most important destruction pathways of $\rm NH_2D$ return $\rm NH_3D^+$. This can be seen by inspection of Figs.\,\ref{fig:ammonia_ratios} and \ref{fig:ammonia_ions} which show the results of the full simulation, and from the schematic presentation of the situation at late times in Fig.~\ref{fig:damm_cycle}. The spin state is conserved in the charge exhange reaction $\dammo \arrow^{\hplus} \dammoplus$. Assuming total scrambling of nuclei in the reaction $\dammoplus + \htwo$, ${\rm o}\dammoplus + {\rm p}\htwo$ gives rise to the intermediate complex ${\rm NH_4}(F_1){\rm D^+}$ which dissociates to o$\dammohplus$ and p$\dammohplus$ in ratio 1:2 (the symmetry of ${\rm H_4}$ is indicated in the brackets, see Table~III of \citealt{Hugo09}). The reaction ${\rm p}\dammoplus + {\rm p}\htwo$ produces only p$\dammohplus$ through the intermediate complex ${\rm NH_4}(E){\rm D^+}$ (only
reaction with p$\htwo$ are considered because the o$\htwo$ abundance is very low at this stage). Using these branching ratios one obtains that the closed cycle $\dammo \arrow^{\hplus} \dammoplus \arrow^{\htwo} \dammohplus \arrow^{\el} \dammo$ should result in o/p-$\dammo\sim2.3$. As discussed above, in the full reaction network this ratio is, however, about 20\% lower.   

\begin{figure*}
\centering
\includegraphics[width=2.0\columnwidth]{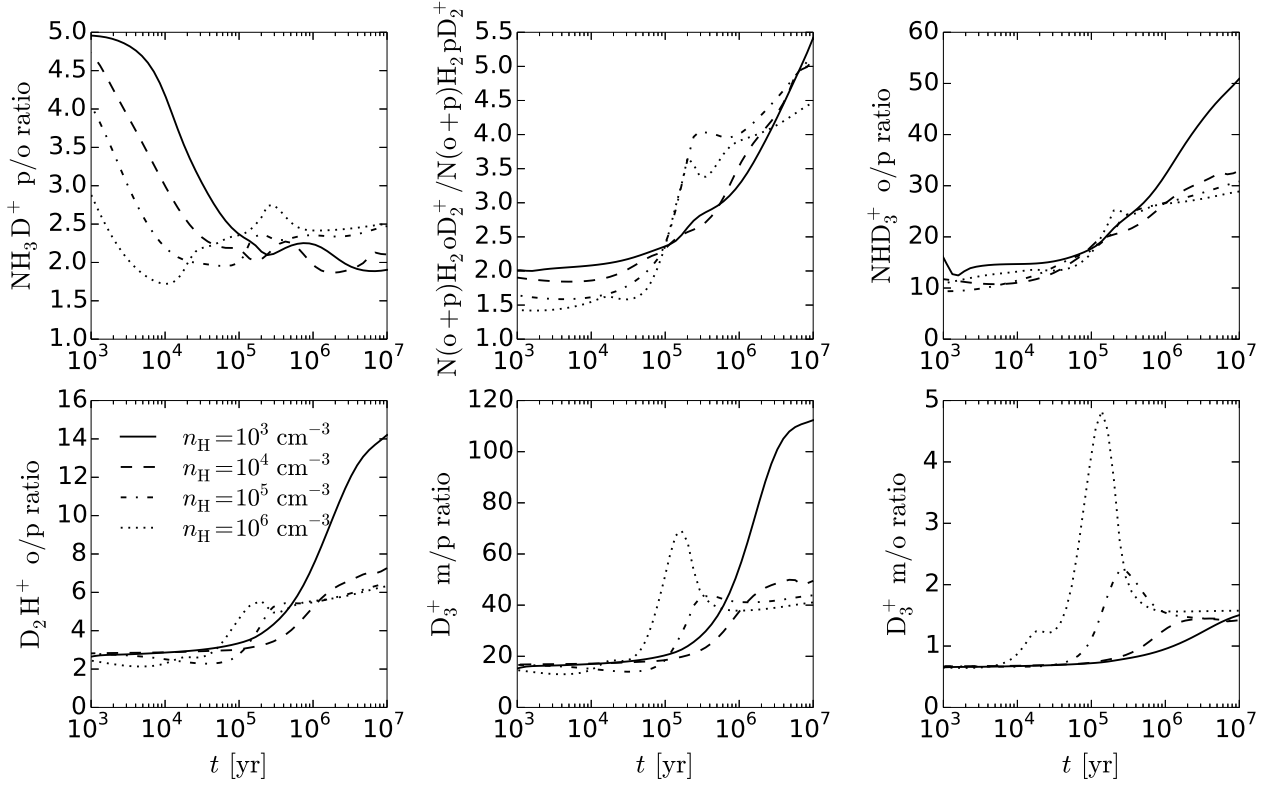}
\caption{Spin-state ratios of the deuterated forms of the ammonium ion, as well as $\rm D_2H^+$ and $\rm D_3^+$, as functions of time at different densities, labeled in the lower left panel.}
\label{fig:ammonia_ions}
\end{figure*}

The doubly and triply deuterated froms of ammonia are primarily formed from the electron recombination of $\ddammohplus$ and $\ndthreehplus$, respectively. (The recombination of the fully deuterated ammonium ion $\rm ND_4^+$ contributes about 10\% to $\ndthree$.)  The cycle $\ddammo \arrow^{\hplus} \dammoplus \arrow^{\htwo} \ddammohplus \arrow^{\el}\ddammo$ preserves the spin state of ${\rm D_2}$. The same is true for $\rm D_3$ in the corresponding cycle of $\ndthree$. $\ddammohplus$ forms mainly from the reaction $\rm D_2H^+ + NH_3 \longrightarrow NH_2D_2^+ + H_2$, where the spin state of $\dtwoh$ is conserved. $\rm NHD_3^+$ receives contribution from both $\rm D_2H^+$ and $\rm D_3^+$ through the reactions $\rm D_3^+ + NH_3
\longrightarrow NHD_3^+ + H_2$ and $\rm D_2H^+ + NH_2D \longrightarrow NHD_3^+ + H_2$. In the first of these reactions, the spin ratios of $\dthree$ are just copied to $\ndthree$. For the second reaction, the branching ratios can be read from Table~IV of \cite{Hugo09}. For example, taking the most common species, the reaction ${\rm o}\dtwoh + {\rm o}\ddammo$ forms an intermediate reaction complex $({\rm NH_2D_4^+})^*$ with $A_1$, $E$, and $F_1$ symmetries of ${\rm D_4}$ in statistical ratios 15:6:15, and, taking these ratios into account, the complex decay into $\ndthreehplus +{\rm H}$ produces $A_1$ (``meta'') and $E$ (``ortho'') symmetries of ${\rm D_3}$ with probabilities 5/9 and 4/9, respectively (the $A_2$ ``para'' symmetry is not formed in this reaction because in ${\rm D_4}\rightarrow {\rm D_3}+{\rm D}$, $A_1\rightarrow A_1$, $E \rightarrow E$, and $F_1 \rightarrow A_1 + E$, Table IV of \citealt{Hugo09}).

What is said above implies that the spin ratios of $\ddammo$ and $\ndthree$ are tied to those of $\dtwoh$ and $\dthree$, and this is also evident from Figs.~\ref{fig:ammonia_ratios} and \ref{fig:ammonia_ions}. The o/p ratio of $\rm NHD_2$ increases constantly until very late times, levelling off only after $10^7$ years in the models shown in Fig.~\ref{fig:ammonia_ratios} (i.e., outside the range shown here). This depends on the $\rm D_2H^+$ o/p ratio which increases strongly as a function of time (Fig.\,\ref{fig:ammonia_ions}). From Fig.\,\ref{fig:ammonia_ratios} it is observed that para-$\rm ND_3$ is clearly the least abundant of
the three spin states of $\rm ND_3$, while the ortho and meta states are equally abundant at late times regardless of the density. This behaviour reflects the spin ratios of $\dthree$ shown in Fig.~\ref{fig:ammonia_ions}.

The statistical values for the spin-state ratios of ortho/para $\rm NH_2D$, ortho/para $\rm NHD_2$ and ortho/meta/para $\rm ND_3$ are 3:1, 2:1, and 16:10:1, respectively. From Fig.\,\ref{fig:ammonia_ratios} it is evident that the various ratios deviate strongly from the statistical values at late times, as was already found for ammonia in S15 (see also Fig.\,\ref{fig:s15test}). The various spin-state ratios are generally rather insensitive to variations in the density. Some density dependence is usually present between $10^5$ and $10^6$ years of chemical evolution, but even in this time interval the difference between the low and high density cases is less than a factor of 2. 

Assuming nuclear spin conservation and that the same selection rules as used in the gas phase can be applied to the H and D addition reactions on grain surfaces, the different spin isomers of ammonia should form on grains according to their nuclear spin statistical weights indicated above. This is what is seen in the beginning of the simulation. Later on, when the gas-phase ammonia starts to accrete, the spin ratios change also on grains, finally mimicking the ratios in the gas.

Finally, we note that for the lowest density considered ($n_{\rm H} = 10^3 \, \rm cm^{-3}$), many of the plotted spin ratios are increasing even at very long timescales, for example the m/p ratio of $\rm ND_3$ which increases by a factor of $\sim 2$ from $10^6$ to $10^7$ yr at this density. We note that this effect is due to the very low density and the spin ratios are expected to stabilize at still longer timescales.

\subsection{Temperature dependence of ammonia D fractionation and spin-state ratios}

\begin{figure*}
\centering
\includegraphics[width=2.0\columnwidth]{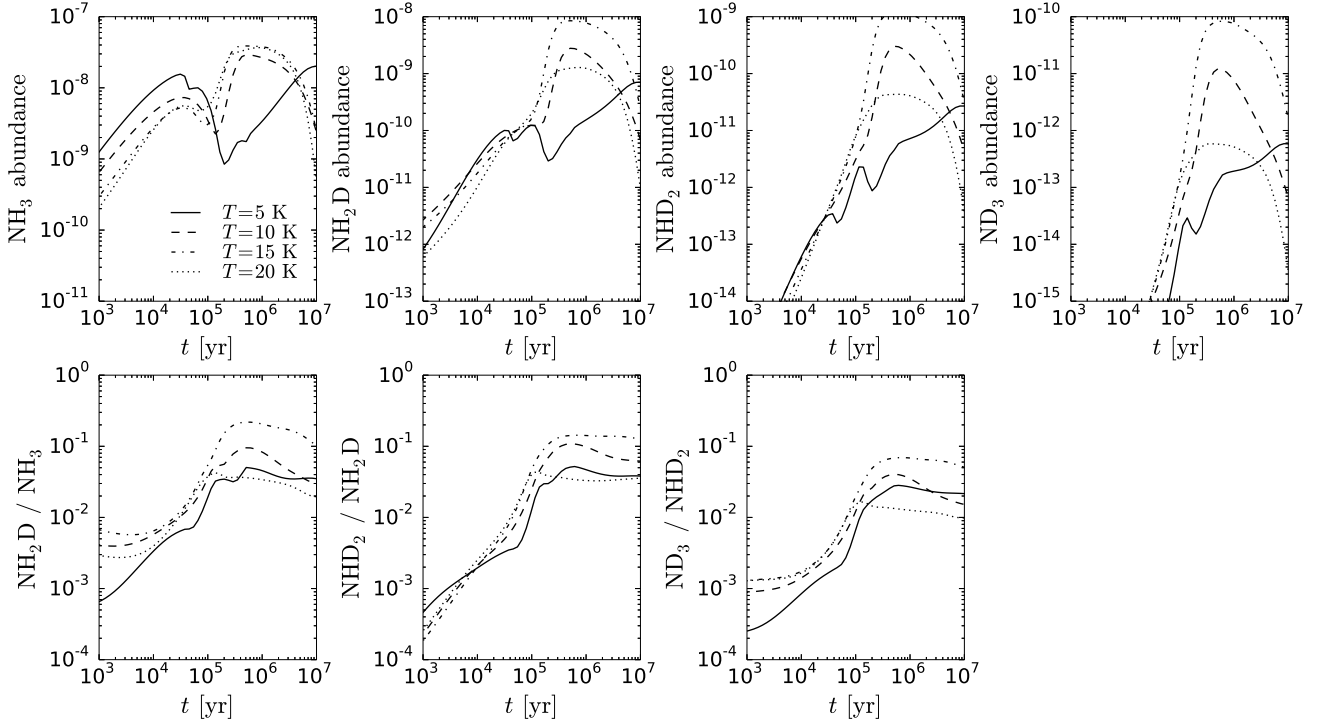}
\caption{{\sl Upper panels:} abundances of ammonia and its deuterated forms as functions of time at different temperatures, labeled in the left-hand panel. {\sl Lower panels:} abundance ratios of ammonia and its deuterated forms as functions of time at different temperatures. In all panels, a constant density of $n_{\rm H} = 10^5\,\rm cm^{-3}$ is adopted.
}
\label{fig:temp_dep}
\end{figure*}

It is well established that deuterium fractionation depends not only on the density, but also on the temperature. In particular, deuteration is expected (in starless cores) to be at its strongest in the temperature range $\sim 10-20\,\rm K$, because on the one hand this range lies below the sublimation temperature of CO, allowing the main driver ion of deuterium chemistry, $\rm H_2D^+$, to transfer its deuteron to other species, and on the other hand deuteration efficiency drops toward low temperatures because of the decreasing accretion rates (see below).

We plot in Fig.\,\ref{fig:temp_dep} the abundances of ammonia and its deuterated forms, and the associated abundance ratios, at different temperatures. Evidently, the deuterium fractionation is strongest around $T = 15$\,K; both the abundances and the deuterium fractionation peak at this temperature. An analysis of the reaction rates shows that if the temperature is reduced from 15\,K to 5\,K, the decreased accretion rates lead to relatively high abundances of atomic H and D in the gas phase, which enable efficient conversion of $\rm D_2H^+$ to $\rm H_2D^+$ through the exchange reaction $\rm D_2H^+ + H \longrightarrow H_2D^+ + D$. This decreases the ammonia deuterium fractionation because $\rm D_2H^+$ is an important part of the production paths of the deuterated forms of ammonia (Fig.\,\ref{fig:damm_cycle}). On the other hand, as the temperature is increased from 15\,K to 20\,K, CO starts to desorb efficiently, hindering deuteration in general.

The spin-state ratios of the ammonia isotopologs at different temperatures are shown in Fig.\,\ref{fig:ammonia_ratios_temp}. It is observed that the ratios change significantly as the temperature is varied, and that the variation is much larger than that induced by changes in the density. A minimum in the ortho/para (and meta/para for $\rm ND_3$) ratios is observed at $T = 10\,\rm K$, except for the ortho/para ratio of $\rm NH_2D$ which has a maximum at this temperature. These are in contrast to the overall deuteration, which peaks at $T = 15\,\rm K$.

As explained in Sect.\,\ref{ss:spinratios}, the spin-state ratios of $\rm ND_3$ are determined by those of $\rm NHD_3^+$, which are in turn determined by $\rm D_2H^+$ and $\rm D_3^+$. Moving from 10\,K to 20\,K, the ortho/para ratio of $\rm H_2$ increases, leading to higher ortho/para $\rm D_2H^+$ and meta/ortho (and meta/para) $\rm D_3^+$ ratios, which ultimately increase the ortho/para and meta/para ratios of $\rm ND_3$ with respect to the values at 10\,K. On the other hand at 5\,K, the $\rm H_2$ ortho/para ratio is low, but the abundances of atomic H and D increase owing to decreased accretion rates, and the proton/deuteron exchange reactions with the $\rm H_3^+$ isotopologs (for example, $\rm D_2H^+ + D \longrightarrow D_3^+ + H$) increase in importance in determining the various spin-state ratios. The spin-state ratios of $\rm NH_2D$ and $\rm NHD_2$ are affected by similar effects with varying temperature.

\begin{figure}
\centering
\includegraphics[width=\columnwidth]{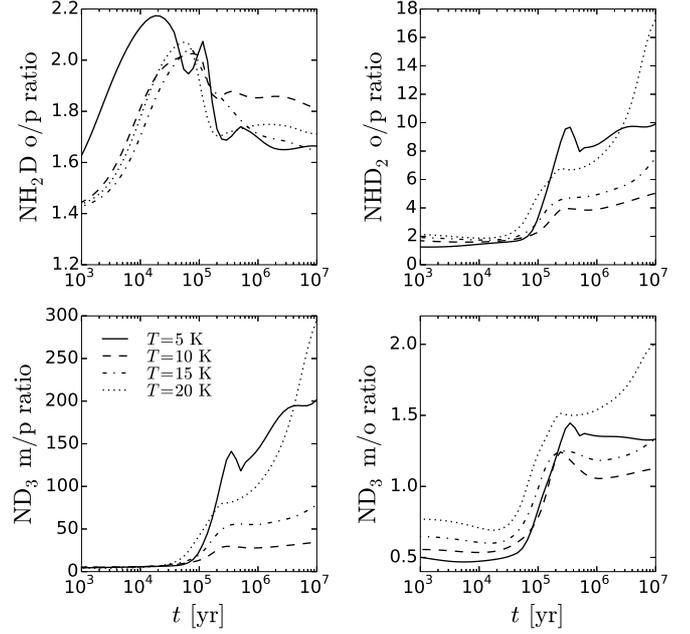}
\caption{Spin-state ratios of the deuterated forms of ammonia as functions of time at different temperatures, labeled in the lower left panel. A constant density of $n_{\rm H} = 10^5\,\rm cm^{-3}$ is adopted.
}
\label{fig:ammonia_ratios_temp}
\end{figure}

\section{Discussion} \label{s:discussion}

\subsection{Radial profiles for deuterated ammonia}

From the above it is clear that both the deuterium fractionation of ammonia and the spin-state ratios depend on time and on the physical parameters (density, temperature). To investigate the implication to observations of deuterated ammonia, we constructed radial abundance profiles for ammonia and its deuterated forms by following the method discussed in \citet{Sipila12} and \citet{Sipila13}, in which chemical calculations are combined with a radiative transfer model to produce self-consistent chemical abundance profiles as functions of density, temperature and time. For these calculations, we adopt as the core model a modified Bonnor-Ebert sphere \citep{Galli02, Keto05, Sipila11} of mass $M = 1.0\,M_{\odot}$. We assumed a visual extinction $A_{\rm V} = 10$\,mag at the edge of the core.

\begin{figure*}
\centering
\includegraphics[width=2.0\columnwidth]{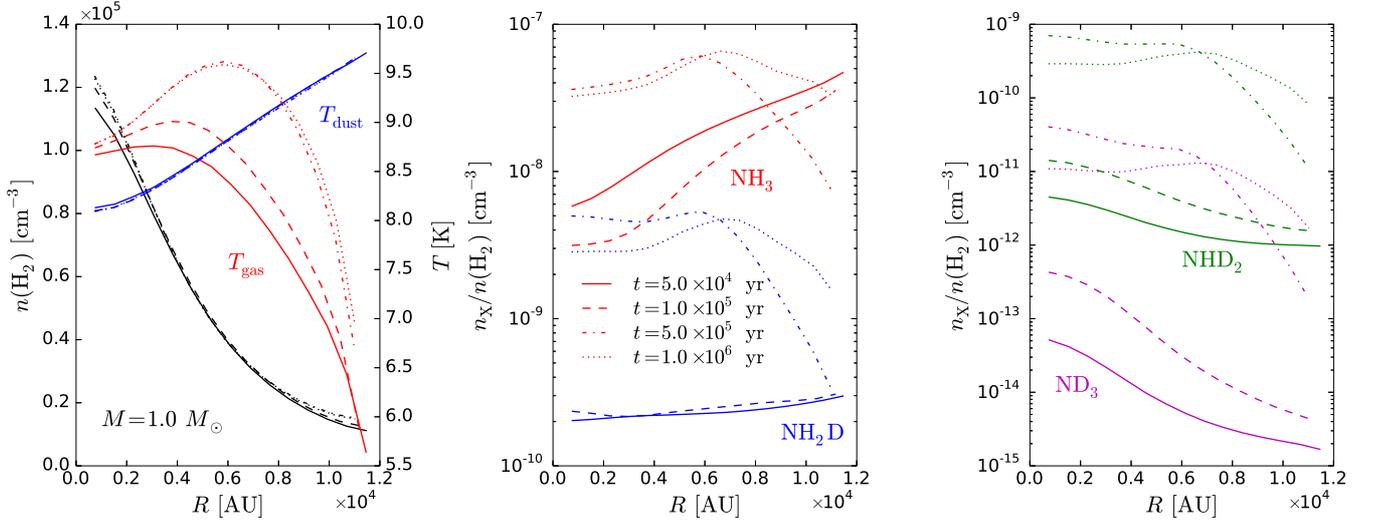}
\caption{{\sl Left-hand panel:} Gas density (black), gas temperature (red) and dust temperature (blue) as functions of distance from the core center at different time steps (indicated in the middle panel). {\sl Middle panel:} Abundances of $\rm NH_3$ and $\rm NH_2D$ as functions of distance from the core center at different time steps. {\sl Right-hand panel:} Abundances of $\rm NHD_2$ and $\rm ND_3$ as functions of distance from the core center at different time steps (same legend as in the other panels). The panels correspond to an MBES of mass $1.0\,M_{\odot}$.
}
\label{fig:MBES}
\end{figure*}

The density and temperature profiles of the model core are shown in the left-hand panel in Fig.\,\ref{fig:MBES}. The density and dust temperature profiles hardly change as functions of time. The density is not high enough even in the core center to equalize the dust and gas temperatures. On the other hand, the gas temperature decreases toward the outer areas because of efficient line cooling by (mainly) CO. As the chemical evolution progresses, CO depletes on grain surfaces, reducing the cooling rates, and the gas temperature rises.

The calculated radial abundance profiles are shown in the middle and right-hand panels in Fig.\,\ref{fig:MBES}. The time steps shown in the figure refer to years of chemical evolution with respect to an initially atomic initial state (except for H and D which are initially in $\rm H_2$ and HD; see S15). The abundances of ammonia and its deuterated isotopologs generally increase as a function of time. Notably, the abundance of $\rm ND_3$ increases by about three orders of magnitude across the core from $t = 5 \times 10^4$\,yr to $t = 1 \times 10^6$\,yr, and the increase in time is stronger for species with more D atoms. The difference in abundance between the core center and edge can be close to two orders of magnitude depending on the species considered and on the time. Depletion slightly decreases the abundances at the core center at the last time step, while the abundances keep increasing at the edge where the density is lower and little depletion is present. An exception to the latter behavior is ammonia, whose abundance has a minimum in the outer core around $t = 5 \times 10^5$\,yr; this kind of behavior is also displayed for example in Fig.\,\ref{fig:temp_dep} and is discussed in Sect.\,\ref{ss:ammonia_abu}.

The results shown here demonstrate on the one hand the importance of accurate knowledge of the physical structure, and on the other hand the necessity of detailed chemical modeling in ascertaining the chemical age of the object in question. However, quantifying the effect of the varying abundance profiles on observable emission requires calculation of the line profiles with a radiative transfer model; this will the subject of a future paper.

\begin{figure}
\centering
\includegraphics[width=\columnwidth]{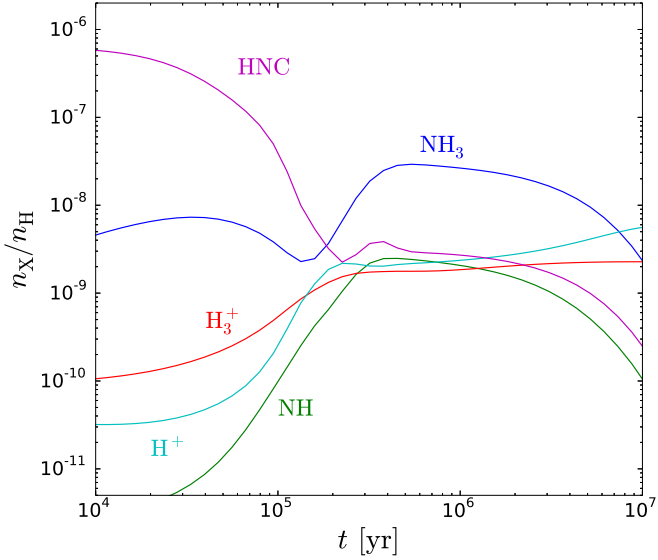}
\caption{Fractional abundances (with respect to $n_{\rm H}$) of selected species (labeled in the plot) in a homogeneous model with $n_{\rm H} = 10^5 \rm \, cm^{-3}$, $T = 10$\,K.
}
\label{fig:ammonia_evolution}
\end{figure}

\subsection{Evolution of the ammonia abundance} \label{ss:ammonia_abu}

Characteristic for the time dependence of the $\rm NH_3$ abundance at high densities, and to a lesser extent for that of $\rm NH_2D$, is a minimum occurring somewhere between $10^5$ and $10^6$ yr, and a rapid recovery after this minimum. This tendency is most clearly visible in Fig.\,\ref{fig:temp_dep}. The behavior is very different from that observed in the case of CO or CS, for example, but resembles the variation of the abundances of HCN and HNC seen in the models of \citet{Loison14}.

The ammonia production is assumed to start slowly, the widely recognized bottle-neck being the slightly endothermic reaction $\rm N^+ + H_2 \longrightarrow NH^+ + H$ which in practice works only with o$\rm H_2$ in cold clouds (e.g., Dislaire et al. 2012; Roueff et al. 2015). Once $\rm NH^+$ is formed it is quickly converted to ammonia through $\rm NH^+ + H_2 \longrightarrow NH_2^+ + H$ etc. In fact, the flow through the bottle-neck is so slow that most of the time the chemical network may find other means to produce $\rm NH^+$. In the present model, $\rm NH^+$ is copiously formed at early times ($\sim 10^4$ yr) from the 
dissociative ionization of HNC by $\rm He^+$, the bottle-neck reaction coming in second. The efficiency of this channel decreases along with the depletion of HNC, and because $\rm NH_3$ accretes onto grains as well, its gas-phase abundance decreases some time before $10^5$ yr, or slightly after that depending on the density.

A strong increase in the ammonia abundance is seen, however, after the minimum, as an indirect consequence of the depletion of CO and neutrals like HCN and OH which are ionized in charge transfer reactions with $\rm H^+$. The disappearance of these molecules enhances greatly the abundances of $\rm H_3^+$ and $\rm H^+$. Firstly, $\rm H_3^+$ converts part of $\rm N_2$ to NH through the reaction $\rm H_3^+ + N_2 \longrightarrow N_2H^+ + H_2$, followed by $\rm N_2H^+ + e^- \longrightarrow N_2 + H/NH + N$. Although less than 10\% of the recombinations lead to NH, this channel is important for ammonia, because the increased abundance of $\rm H^+$ directs NH to the ammonia track through the charge transfer reaction $\rm H^+ + NH \longrightarrow NH^+ + H$. Likewise, all the recombination products of $\rm NH_3^+ + e^-$ and $\rm NH_4^+ + e^-$ (NH, $\rm NH_2$, $\rm NH_3$) are efficiently returned to the ammonia cycle by $\rm H^+$. In the present model, ammonia passes several times the cycle $\ammo \arrow^{\hplus} \ammoplus \arrow^{\htwo} \ammohplus \arrow^{\el} \ammo$ described in Sect.\,\ref{ss:ammoniadfrac} before colliding with a dust grain. In Fig.\,\ref{fig:ammonia_evolution} we show the fractional abundances of $\rm NH_3$, NH, $\rm H^+$, $\rm H_3^+$, and HNC as functions of time for a homogeneous model with $n_{\rm H} = 10^5 \rm \, cm^{-3}$, $T = 10$\,K. In this model, $\rm NH_3$ and NH are closely correlated after $10^5$ yr and follow the gradual decrease of the $\rm N_2$ abundance (not shown) at very late times.

\subsection{Comparison to \citet{Roueff05}}\label{ss:roueffcomparison}

\citet{Roueff05} presented observations of $\rm NH_2D$, $\rm NHD_2$, and $\rm ND_3$ toward prestellar cores, and discussed the interpretation of the observations in the context of a gas-phase chemical model. Their model is to our knowledge the most directly comparable to our own, in terms of the chemistry of deuterated ammonia.

Our model includes time-dependent depletion\footnote{In our model, nitrogen and deuterium deplete as well, unlike in the model of \citet{Roueff05}.}, and a chemical steady-state is not reached in the time intervals considered here. However, comparing the values given in Table~9 in \citet{Roueff05} with our results near the deuterium peak for each density (Fig.\,\ref{fig:ammonia_abus}), we find that the agreement between the models is very good, despite the fact that we are using a gas-grain model with different initial abundances and a different value for the cosmic ray ionization rate (here $1.3 \times 10^{-17}\,\rm s^{-1}$; $2 \times 10^{-17}\,\rm s^{-1}$ in Roueff et al.). It should be noted, however, that in our model, depletion occurs also at a density of $10^4\,\rm cm^{-3}$, whereas in the Roueff et al. model no depletion is assumed at this density.

\begin{figure}
\centering
\includegraphics[width=\columnwidth]{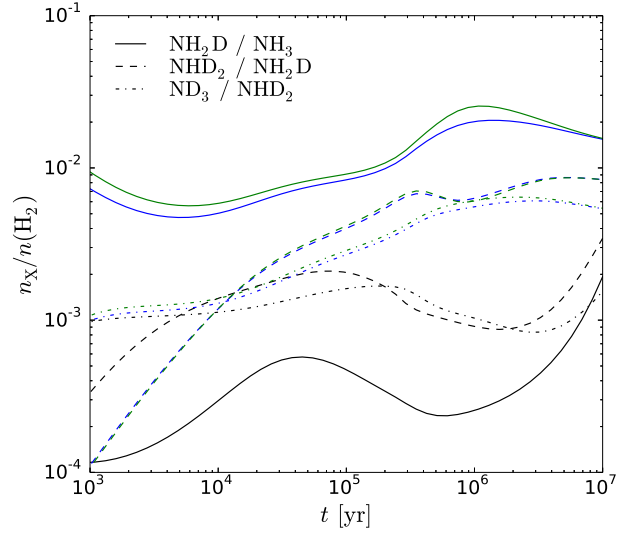}
\caption{Abundance ratios of ammonia and its deuterated forms (labeled in the plot) as functions of time in a gas-phase chemical model, for different initial values of the $\rm H_2$ ortho/para ratio: 3 (black lines), $10^{-3}$ (blue), or $10^{-5}$ (green).
}
\label{fig:ammonia_gasphase}
\end{figure}

As an additional check, we ran another calculation using a gas-phase model (i.e., no depletion) at $n({\rm H_2}) = 10^4\,\rm cm^{-3}$, for three different values of the initial $\rm H_2$ ortho/para ratio. The resulting abundance ratios for ammonia and its deuterated forms are shown in Fig.\,\ref{fig:ammonia_gasphase}. The gas-phase model predicts a lower deuterium fractionation than the gas-grain model, even for very low initial $\rm H_2$ ortho/para ratios. The main reason for this is that the high abundance of CO hinders the formation of $\rm H_2D^+$, $\rm D_2H^+$ and $\rm D_3^+$, which are in our model the principal agents in the formation of deuterated ammonia.

\citet{Roueff05} assumed that the release of hydrogen is favored (in a 2:1 ratio) over the release of deuterium in the electron recombinations of the various deuterated ammonium ions, whereas in this work the H/D branching ratios for the recombination reactions are statistical. We tested the effect of assuming a 2:1 ratio for the recombination reactions and found that this has only a marginal effect on the abundances of ammonia and its isotopologs. We note that we adopt the $\rm NH_3^+ + H_2 \rightarrow NH_4^+ + H$ reaction rates from \citet{LeGal14} which correspond to the values used by \citeauthor{Roueff05} (\citeyear{Roueff05}; see also \citealt{Roueff15}). We tested the new coefficients given by \citet{Roueff15} for these reactions and found that they have only a minor effect on our results.

The differences between our gas-phase model and that of \citet{Roueff05} can perhaps be explained by the adoption of different rate coefficients for the $\rm H_3^+ + H_2$ reacting system, since the contribution of the $\rm H_3^+$ isotopologs to the chemistry of deuterated ammonia is significant as discussed earlier in this paper. Finally we reiterate that the depletion present in our gas-grain model even at lower densities enhances deuteration and brings the results close to the gas-phase model of \citet{Roueff05}.

\section{Conclusions}

We calculated branching ratios for reactions involving multiply deuterated species using group theory. This method produces identical branching ratios for reactions without deuterium, which can be derived with angular momentum algebra \citep{Oka04}. Our new model allows us to track the spin states of any species that includes multiple protons and/or deuterons. However, in this paper, we concentrated on the spin-state chemistry of deuterated ammonia.

We found that the deuterium fractionation of ammonia is high especially at high density: the $\rm NH_2D / NH_3$, $\rm NHD_2 / NH_2D$, and $\rm ND_3 / NHD_2$ all peak at $\sim$ 10\% for $n_{\rm H} = 10^6 \, \rm cm^{-3}$. The high abundance ratios are in line with earlier gas-phase models of \citet{Roueff05}. The various spin-state ratios were found to deviate strongly from the statistical values especially at late times, but on the other hand the ratios appear rather insensitive to variations in the density. We also found that the abundances of ammonia and its isotopologs and the associated spin-state ratios are highly dependent on the temperature; the meta/para ratio of $\rm ND_3$ for example can vary by a factor of $\sim 5$ in the temperature range $5 - 20$\,K. 

The abundances of $\rm NH_2D$, $\rm NHD_2$, and $\rm ND_3$, and their spin-state ratios in the gas phase, are determined by competing ion-molecular reactions, the rates of which depend strongly on the kinetic temperature. The deuterium fractionation is most efficient at 15 K. The temperature dependences of the spin-state ratios are less straightforward: the para species of $\rm NHD_2$ and $\rm ND_3$ reach their abundance maxima at 10 K, wheras para-$\rm NH_2D$ has a minimum at this temperature. 
The reactions leading to deuterated ammonia involve $\rm H_2D^+$, $\rm D_2H^+$, and $\rm D_3^+$, and benefit from the disappearance of CO which occurs faster at high densities. Therefore, for objects with density/temperature gradients, such as starless and prestellar cores, significant variations as functions of radius and time can be expected in the abundances and the spin-state ratios of $\rm NH_2D$, $\rm NHD_2$, and $\rm ND_3$. Sensitive observations of these molecules, combined with chemical and radiative transfer modelling can provide stringent constraints on the physical and chemical structures, and the evolutionary stages of molecular clouds.

The present model does not include corrections to rate coefficients resulting from differences in the zero-point energies of deuterated and undeuterated species, except for the $\rm H_3^+ + H_2$ reacting system given by \citet{Hugo09} and the $\rm N^+ + H_2$ reaction \citep{Marquette88}. Moreover, the use of pure nuclear spin statistical weights is based on the implicit assumption that the exothermicity of the reaction is sufficiently large so that Pauli allowed rotational levels of each product species can be populated. For the principal reactions contributing to the interstellar ammonia, this is likely to be true (see also discussion in \citealt{Rist13}). However, it is clear that the present model is still incomplete, and an in-depth study of energetics is warranted. 

\begin{acknowledgements}

We thank the anonymous referee for a careful reading of the manuscript, and Laurent Wiesenfeld for helpful discussions. O.S. and P.C. acknowledge the financial support of the European Research Council (ERC; project PALs 320620) 

\end{acknowledgements}

\bibliographystyle{aa}
\bibliography{refs.bib}

\appendix

\def\Harpoons{\mathop{\rightleftharpoons}\limits}

\section{Calculation of nuclear spin branching ratios using symmetry rules}

The method used in the derivation of the nuclear spin branching ratios is demonstrated by three examples, all dealing with varieties of the reaction $\rm H_3^+ + NH_3 \rightarrow$ $(\rm NH_6^+)^*\rightarrow$ $\rm NH_4^+ + H_2$. The method is described in \cite{Quack77} and in \cite{Park07}. The collision partners are classified, or labeled, according to the irreducible representations of an appropriate permutation group. These labels tell how the symmetrized nuclear spin functions belonging to this species transform under operations of the group, like the transposition of two identical nuclei and the permutation of $n$ identical nuclei (e.g., \citealt{BunkerJensen06}). The branching ratios are obtained from correlations between the group of the intermediate reaction complex and the direct product groups representing the reactants and products \citep{Quack77}. The statistical weight of each channel corresponds to the number of nuclear spin functions belonging to that symmetry species.

\subsection{Three identical nuclei}

As the first example we consider the reaction $\rm D_2H^+ + NH_2D \rightarrow$ $(\rm NH_3D_3^+)^* \rightarrow$ $\rm NHD_3^+ + H_2$. The H and D symmetries are treated separately. The same result would be obtained by considering the mixed system in appropriate direct product groups. From the hydrogen point of view, $\rm H_2$ and $\rm H$ form $\rm H_3$ which dissociates back to $\rm H_2 + H$. On the deuterium side, the reaction forms $\rm D_3$ from $\rm D_2 + D$. In the present model we assume that the branching ratios are equal to the pure nuclear spin symmetry induction and subduction statistical weights, which for these reactions can be read from Tables III and IV of \cite{Hugo09}. We reproduce the appropriate tables in Table~\ref{table:H3_and_D3}. 

\begin{table}
\caption{Statistical nuclear spin branching ratios for reactions 
$\rm H_2 + H \Harpoons {\rm H_3}$ and 
$\rm D_2 + D \Harpoons {\rm D_3}$.}
\begin{minipage}{4.3cm}
\begin{tabular}{c|cc|c}
{${\rm H_3}$} & \multicolumn{2}{|c|}{$\rm H_2 + H$}& \\
     &  $A\otimes A$ & $B\otimes A$  & $\sum$ \\ \hline
4\,$A_1$  &  4   &  0     &   4    \\
0\,$A_2$  &  0   &  0    &   0    \\ 
2\,$E$    &  2   &  2    &   4    \\ \hline
{$\sum$} & 6   &  2    &   8    \\  
\end{tabular}
\end{minipage}
\begin{minipage}{4cm}
\begin{tabular}{c|cc|c}
{${\rm D_3}$} & \multicolumn{2}{|c|}{$\rm D_2 + D$} & \\
    &  $A\otimes A$ & $B\otimes A$  & $\sum$ \\ \hline
10\,$A_1$  &  10   &  0     &   10    \\
1\,$A_2$  &  0   &  1    &   1    \\ 
8\,$E$    &  8   &  8    &   16    \\ \hline
{$\sum$} & 18   &  9    &   27    \\  
\end{tabular}
\end{minipage}
\label{table:H3_and_D3}
\end{table}

The numbers in the leftmost columns of Table~\ref{table:H3_and_D3} are the frequencies of the irreducible representations $A_1$, $A_2$, and $E$, in the representations of the permutation group $S_3$ generated by the nuclear spin functions of ${\rm H_3^+}$ and ${\rm D_3^+}$. This means that the symmetry representations of ${\rm H_3}$ and ${\rm D_3}$ in $S_3$ can be reduced into the following direct sums (denoted by $\oplus$):
\begin{equation}
\Gamma^{\rm n.s.}({\rm H_3}) = 4\,A_1 \oplus 2\,E  \; \; \; , \; \; \; 
\Gamma^{\rm n.s.}({\rm D_3}) = 10\,A_1\oplus 1\,A_2 \oplus 8\,E
\end{equation}
(e.g., \citealt{BunkerJensen06}; \citealt{Hugo09}). According to Mulliken's notation used here, one-dimensional representations are labelled either $A$ or $B$, two dimensional are labelled $E$, three-dimensional $F$, four-dimensional $G$, etc.  The total nuclear spin statistical weight of each symmetry species can be read from these decompositions by multiplying the frequency by the dimension of the representation, $f(\Gamma_i)\times {\rm dim}(\Gamma_i)$. The Greek appellations, {\sl para}, {\sl meta}, and {\sl ortho}, are used to indicate these statistical weights for the simplest molecules, so that ``ortho'' has the highest statistical weight. The ``para'' ($E$, $I=1/2$) and ``ortho'' ($A_1$, $I=3/2$) species of ${\rm H_3^+}$ have, however, equal statistical weights. ${\rm H_3^+}$ has no nuclear spin functions of species $A_2$ which is antisymmetric with respect to interchange of two nuclei\footnote{The fulfilment of the symmetrization postulate for each nuclear spin species is taken care of by the complete internal wavefunction consisting of the nuclear spin, ro-vibrational and electronic parts.}. ${\rm H_2}$ and ${\rm D_2}$ can have the symmetries $A$ or $B$, and the single nuclei H and D have the $A$ symmetry. The number in the bottom right corner of each table gives the total number of linearly independent nuclear spin functions of the molecule, $(2I+1)^3$, where $I=1/2$ for ${\rm H_3}$ and $I=1$ for ${\rm D_3}$. The branching ratios in Table~\ref{table:H3_and_D3} are obtained by multiplying the correlation table between $S_3$ and the product group $S_2 \otimes S_1$ by appropriate frequencies. This correlation table is given by
 
\begin{center} 
\begin{tabular}{c|cc} 
$S_3$ & \multicolumn{2}{|c}{$S_2 \otimes S_1$} \\ 
 & $A\otimes A$ & $B\otimes A$ \\ \hline 
$A_1$ & 1 & 0 \\ 
$A_2$ & 0 & 1 \\ 
E & 1 & 1 
\end{tabular}
\end{center} 
 
The correlation table $S_3 \leftrightarrow S_2\otimes S_1$ presented above is derived using the standard methods of group theory, i.e., by forming the character table of direct product group $S_2 \otimes S_1$, and the character table for the subduced representations of $S_3$ on $S_2 \otimes S_1$, $S_3\downarrow S_2\otimes S_1$. The latter is obtained from that of $S_3$ by stripping it from operations which are unfeasible in the direct product group $S_2\otimes S_1$, i.e., the class of cyclic permutations of three nuclei, $\left\{(123),(132)\right\}$. Finally, the orthogonality relation between the characters of irreducible representations is used.

The Tables~\ref{table:H3_and_D3} can be read either from left to right ("subduction") or top down ("induction"). For example, one can see that the $A_1$ species of ${\rm H_3}$ and ${\rm D_3}$ dissociate exclusively to the A symmetry of ${\rm H_2}$ or ${\rm D_2}$, whereas the $E$ species dissociate to both $A$ and $B$ symmetries (``ortho" and``para", respectively) of ${\rm H_2}$ or ${\rm D_2}$ with equal probability. In the reaction ${\rm oD_2} + {\rm D}$ ($A\otimes A$), the $A_1$ (``meta") and $E$ (``ortho") species of ${\rm D_3}$ are formed with the probabilities 10/18 and 8/18, respectively, whereas in the reaction ${\rm pD_2} + {\rm D}$ ($B\otimes A$), the $A_2$ (``para") and $E$ (``ortho") species of ${\rm D_3}$ are formed with the probabilities 1/9 and 8/9. These branching ratios determine relative abundances of ${\rm NHoD_3^+}$, ${\rm NHmD_3^+}$, and ${\rm NHpD_3^+}$ resulting from the reaction ${\rm D_2H^+} + {\rm NH_2D}$. 

\subsection{Six identical nuclei}

In order to determine the nuclear spin branching ratios of the reactions ${\rm H_3^+} + {\rm NH_3} \rightarrow$ $({\rm NH_6^+})^*
\rightarrow$ ${\rm NH_4^+} + {\rm H_2}$ and ${\rm D_3^+} + {\rm ND_3} \rightarrow$ $({\rm ND_6^+})^* \rightarrow$ ${\rm ND_4^+} + {\rm D_2}$ one needs to determine the nuclear spin symmetry species of ${\rm H_6}$ and ${\rm D_6}$ in the permutation group $S_6$. We calculate first the characters, $\chi(R)$, of the representations generated by the nuclear spin functions of ${\rm H_6}$ and ${\rm D_6}$ under the operations $R$ of $S_6$. \cite{Crabtree13} give a convenient formula for this:
\begin{equation}
\chi(R) = (2I+1)^{n-x} \; , 
\label{eq:redchar}
\end{equation}
where $I$ is the nuclear spin ($1/2$ for H and $1$ for D), $n$ is the number of identical nuclei (here 6), and $x$ is the number of transpositions or pairwise permutations which correspond to the operation $R$.  The number of interchanges corresponding to a cyclic permutation of $k$ elements is $k-1$, and so $x=k-1$. For example, $x=2$ for the ternary permutation $(123)$. The character table of $S_6$ is easily obtainable using \cite{GAP4}. By applying the orthogonality relation of the characters of irreducible representations, we derive the following decompositions for the symmetry representations of ${\rm H_6}$ and ${\rm D_6}$:
\begin{equation}
\Gamma^{\rm n.s.}({\rm H_6}) =  7\,A_1 \oplus 5\, H_1 \oplus 1\, H_4 \oplus 3\,L_1   \; , 
\end{equation}
\begin{equation}
\Gamma^{\rm n.s.}({\rm D_6}) = 28\, A_1 \oplus 35\,H_1 \oplus 1\,H_3
  \oplus 10\,H_4 \oplus 27\,L_1 \oplus 10\,M_1 \oplus 8\,S  \,  .
 \end{equation}
The correlation tables $S_6 \leftrightarrow S_3\otimes S_3$ and $S_6 \leftrightarrow S_4\otimes S_2$ needed for the induction and subduction statistics are derived as outlined in the previous example. Finally, the branching ratio tables listed in Tables~\ref{table:H6toH3+H3}, \ref{table:H6toH4+H2}, \ref{table:D6toD3+D3}, and \ref{table:D6toD4+D2} are obtained by multiplying the correlation tables by the frequencies of irreducible representations appearing in the decompositions given above. In addition, the columns of the correlation table are multiplied by the dimensions of the subduced representations in order to make the weights correspond to the total number of nuclear spin functions belonging to each representation. In the previous example, all products had one-dimensional representations. The correct normalisation of the weights can be checked by inspecting the last rows and columns of the tables. In these tables, species with zero frequencies are not listed.

\begin{table}
\caption{Statistical nuclear spin branching ratios of the reaction ${\rm H_6} \Harpoons {\rm H_3} + {\rm H_3}$.}
\label{table:H6toH3+H3}

\begin{tabular}{c|cccc|c}
${\rm H_6}$ & \multicolumn{4}{|c|}{${\rm H_3} + {\rm H_3}$} & \\ 
      & $A_1,A_1$ & $A_1,E$ & $E,A_1$ & $E,E$ & $\sum$ \\ \hline
 $7\,A_1$ &      7&        0&       0&         0&     7\\
 $5\,H_1$ &      5&       10&      10&         0&    25\\
 $1\,H_4$ &      1&        0&       0&         4&     5\\
 $3\,L_1$ &      3&        6&       6&        12&    27\\ \hline
  $\sum$&     16&       16&      16&        16&    64\\
\end{tabular}

\end{table}

\begin{table}
\caption{Statistical nuclear spin branching ratios of the reaction ${\rm H_6} \Harpoons {\rm H_4} + {\rm H_2}$.}
\label{table:H6toH4+H2}
\begin{tabular}{c|cccccc|c}
${\rm H_6}$ & \multicolumn{6}{|c|}{${\rm H_4} + {\rm H_2}$} & \\ 
      & $A_1,A$ & $A_1,B$ & $E,A$ &  $E,B$ & $F_1,A$ & 
        $F_1,B$ & $\sum$ \\ \hline
 $7\,A_1$  &    7 &    0 &     0 &    0 &    0 &    0 &      7\\
 $5\,H_1$  &    5 &    5 &     0 &    0 &   15 &    0 &     25\\
 $1\,H_4$  &    0 &    0 &     0 &    2 &    3 &    0 &      5\\
 $3\,L_1$  &    3 &    0 &     6 &    0 &    9 &    9 &     27\\ \hline
  $\sum$ &   15 &    5 &     6 &    2 &   27 &    9 &     64\\
\end{tabular}
\end{table}

\begin{table*}
\caption{Statistical nuclear spin branching ratios of the reaction ${\rm D_6} \Harpoons {\rm D_3} + {\rm D_3}$.}
\label{table:D6toD3+D3}

\begin{tabular}{c|ccccccccc|c}
${\rm D_6}$ & \multicolumn{9}{|c|}{${\rm D_3} + {\rm D_3}$} & \\ 
      & $A_1,A_1$ & $A_1,A_2$ & $A_1,E$ & $A_2,A_1$ &  $A_2,A_2$ &  $A_2,E$ & $E,A_1$ & 
        $E,A_2$ & $E,E$ & $\sum$ \\ \hline
 $28\,A_1$ &  28&     0&     0&     0&     0&     0&     0&     0&     0&    28\\
 $35\,H_1$ &  35&     0&    70&     0&     0&     0&    70&     0&     0&   175\\
 $1\,H_3$  &   0&     0&     0&     0&     1&     0&     0&     0&     4&     5\\
 $10\,H_4$ &  10&     0&     0&     0&     0&     0&     0&     0&    40&    50\\
 $27\,L_1$ &  27&     0&    54&     0&     0&     0&    54&     0&   108&   243\\
 $10\,M_1$ &   0&    10&    20&    10&     0&     0&    20&     0&    40&   100\\
 $8\,S$    &   0&     0&    16&     0&     0&    16&    16&    16&    64&   128\\ \hline
  $\sum$ & 100&    10&   160&    10&     1&    16&   160&    16&   256&   729\\
\end{tabular}

\end{table*}

\begin{table*}
\caption{Statistical nuclear spin branching ratios of the reaction ${\rm D_6} \Harpoons {\rm D_4} + {\rm D_2}$.}
\label{table:D6toD4+D2}
\begin{tabular}{c|cccccccc|c}
${\rm D_6}$ & \multicolumn{8}{|c|}{${\rm D_4} + {\rm D_2}$} & \\ 
      & $A_1,A$ & $A_1,B$ &  $E,A$ &  $E,B$ & $F_1,A$ & 
        $F_1,B$ & $F_2,A$ & $F_2,B$ & $\sum$ \\ \hline
 $28\,A_1$ &    28  &    0  &   0  &   0 &    0  &   0 &    0 &    0 &   28\\
 $35\,H_1$ &    35  &  35  &   0  &   0 &  105  &   0 &    0 &    0 &  175\\
 $1\,H_3$   &      0  &    0  &   2  &   0 &    0  &   0 &    0 &    3 &    5\\
 $10\,H_4$  &     0  &    0  &   0  &  20 &   30  &   0 &    0 &    0 &   50\\
 $27\,L_1$  &    27  &    0  &  54  &   0 &   81  &  81 &    0 &    0 &  243\\
 $10\,M_1$ &      0  &  10  &   0  &   0 &   30  &  30 &   30 &    0 &  100\\
 $8\,S$       &      0  &     0  &  16  &  16 &   24  &  24 &   24 &   24 &  128\\ \hline
$\sum$       &    90  &  45  &   72  &  36 &  270  & 135 &   54 &   27 &  729\\
\end{tabular}
\end{table*}

The use of the branching ratio tables is illustrated in Figs.~\ref{figure:H6fromA1+E} and \ref{figure:D6fromA1+A1}. Fig.~\ref{figure:H6fromA1+E} gives the outcome of the reaction ${\rm H_3^+}+{\rm NH_3}$ assuming that one of the molecules is of the $A_1$ (``ortho'') symmetry and the other one is $E$ (``para''). The branching probabilities indicated next to the arrows are obtained from Tables~\ref{table:H6toH3+H3} (reading it top down) and from Table~\ref{table:H6toH4+H2} (reading it from left to right). After
combining the probabilities of pathways leading to different symmetry species of ${\rm NH_4^+}$ one finds that the $F_1$, $A_1$, and $E$ species are formed in ratios $15:7:2$ in this reaction. Fig.~\ref{figure:D6fromA1+A1} describes the branching of the reaction ${\rm D_3^+}+{\rm ND_3}$ assuming that both species are $A_1$ (``meta''). The branching ratios are obtained from Tables~\ref{table:D6toD3+D3} and \ref{table:D6toD4+D2}. This reaction produces the $F_1$, $A_1$, and $E$ species of ${\rm ND_4^+}$ in ratios $9:9:2$.

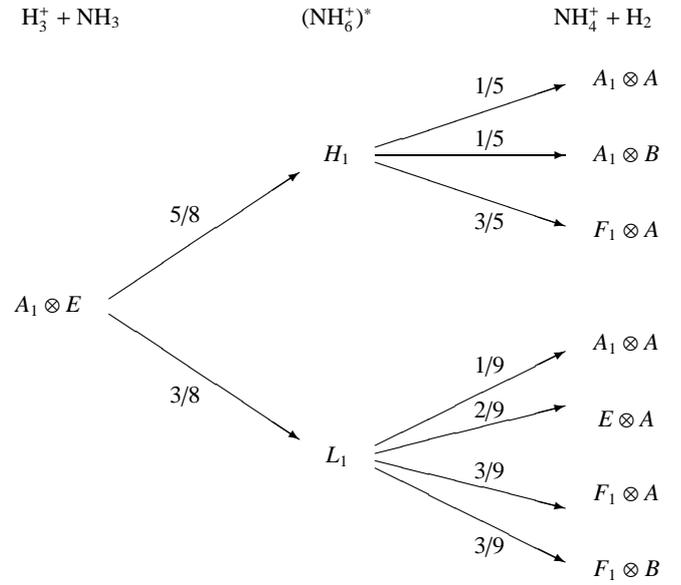
\begin{figure}[hbt]
\centering
\unitlength=1.0mm
\begin{picture}(80,80)(0,0)

\color{black} 

\put(5,78){\makebox(0,0){${\rm H_3^+} + {\rm NH_3}$}}

\put(40,78){\makebox(0,0){$({\rm NH_6^+})^*$}}

\put(75,78){\makebox(0,0){${\rm NH_4^+} + {\rm H_2}$}}

\put(2,40){\makebox(0,0){$A_1\otimes E$}}

\put(40,60){\makebox(0,0){$H_1$}}
\put(40,20){\makebox(0,0){$L_1$}}

\put(78,70){\makebox(0,0){$A_1\otimes A$}}
\put(78,60){\makebox(0,0){$A_1\otimes B$}}
\put(78,50){\makebox(0,0){$F_1\otimes A$}}

\put(78,35){\makebox(0,0){$A_1\otimes A$}}
\put(78,25){\makebox(0,0){$E\otimes A$}}
\put(78,15){\makebox(0,0){$F_1\otimes A$}}
\put(78,5){\makebox(0,0){$F_1\otimes B$}}

\put(10,41){\vector(3,2){25}}
\put(10,39){\vector(3,-2){25}}
\put(20,52){\makebox(0,0){5/8}}
\put(20,28){\makebox(0,0){3/8}}

\put(45,61){\vector(3,1){25}}
\put(45,60){\vector(3,0){25}}
\put(45,59){\vector(3,-1){25}}

\put(60,69){\makebox(0,0){1/5}}
\put(60,62){\makebox(0,0){1/5}}
\put(60,51){\makebox(0,0){3/5}}

\put(45,21.5){\vector(2,1){25}}
\put(45,20.5){\vector(4,1){25}}
\put(45,19.5){\vector(4,-1){25}}
\put(45,18.5){\vector(2,-1){25}}

\put(60,32){\makebox(0,0){1/9}}
\put(60,26){\makebox(0,0){2/9}}
\put(60,18){\makebox(0,0){3/9}}
\put(60,8){\makebox(0,0){3/9}}

\end{picture}

\caption{Branching ratios of the reaction ${\rm H_3^+} + {\rm NH_3}
  \rightarrow ({\rm NH_6^+})^* \rightarrow {\rm NH_4^+} + {\rm H_2}$
  when the reactants are of the $A_1$ (``ortho'') and $E$ (``para'')
  symmetries.}
      \label{figure:H6fromA1+E}
\end{figure}

\begin{figure}[hbt]
\centering
\unitlength=1.0mm
\begin{picture}(80,100)(0,0)

\color{black} 

\put(5,100){\makebox(0,0){${\rm D_3^+} + {\rm ND_3}$}}

\put(40,100){\makebox(0,0){$({\rm ND_6^+})^*$}}

\put(75,100){\makebox(0,0){${\rm ND_4^+} + {\rm D_2}$}}

\put(2,57.5){\makebox(0,0){$A_1\otimes A_1$}}

\put(40,92){\makebox(0,0){$A_1$}}
\put(40,70){\makebox(0,0){$H_1$}}
\put(40,45){\makebox(0,0){$H_4$}}
\put(40,15){\makebox(0,0){$L_1$}}

\put(78,92){\makebox(0,0){$A_1\otimes A$}}

\put(78,80){\makebox(0,0){$A_1\otimes A$}}
\put(78,70){\makebox(0,0){$A_1\otimes B$}}
\put(78,60){\makebox(0,0){$F_1\otimes A$}}

\put(78,50){\makebox(0,0){$E\otimes B$}}
\put(78,40){\makebox(0,0){$F_1\otimes A$}}

\put(78,30){\makebox(0,0){$A_1\otimes A$}}
\put(78,20){\makebox(0,0){$E\otimes A$}}
\put(78,10){\makebox(0,0){$F_1\otimes A$}}
\put(78,0){\makebox(0,0){$F_1\otimes B$}}

\put(10,59){\vector(3,4){25}}
\put(10,58){\vector(2,1){25}}
\put(10,57){\vector(2,-1){25}}
\put(10,56){\vector(2,-3){25}}

\put(45,92){\vector(1,0){25}}

\put(45,71){\vector(3,1){25}}
\put(45,70){\vector(3,0){25}}
\put(45,69){\vector(3,-1){25}}

\put(45,45.5){\vector(4,1){25}}
\put(45,44.5){\vector(4,-1){25}}

\put(45,16.5){\vector(2,1){25}}
\put(45,15.5){\vector(4,1){25}}
\put(45,14.5){\vector(4,-1){25}}
\put(45,13.5){\vector(2,-1){25}}

\put(20,81){\makebox(0,0){28/100}}
\put(24,69){\makebox(0,0){35/100}}
\put(24,45){\makebox(0,0){10/100}}
\put(20,31){\makebox(0,0){27/100}}

\put(60,94){\makebox(0,0){1/1}}

\put(60,79){\makebox(0,0){1/5}}
\put(60,72){\makebox(0,0){1/5}}
\put(60,61){\makebox(0,0){3/5}}

\put(60,51){\makebox(0,0){2/5}}
\put(60,38){\makebox(0,0){3/5}}

\put(60,26){\makebox(0,0){1/9}}
\put(60,21){\makebox(0,0){2/9}}
\put(60,13){\makebox(0,0){3/9}}
\put(60,3){\makebox(0,0){3/9}}

\end{picture}

\caption{Branching ratios of the reaction ${\rm D_3^+} + {\rm ND_3}
  \rightarrow ({\rm ND_6^+})^* \rightarrow {\rm ND_4^+} + {\rm D_2}$
  when both reactants are of the $A_1$ (``meta'') symmetry.}
      \label{figure:D6fromA1+A1}
\end{figure}
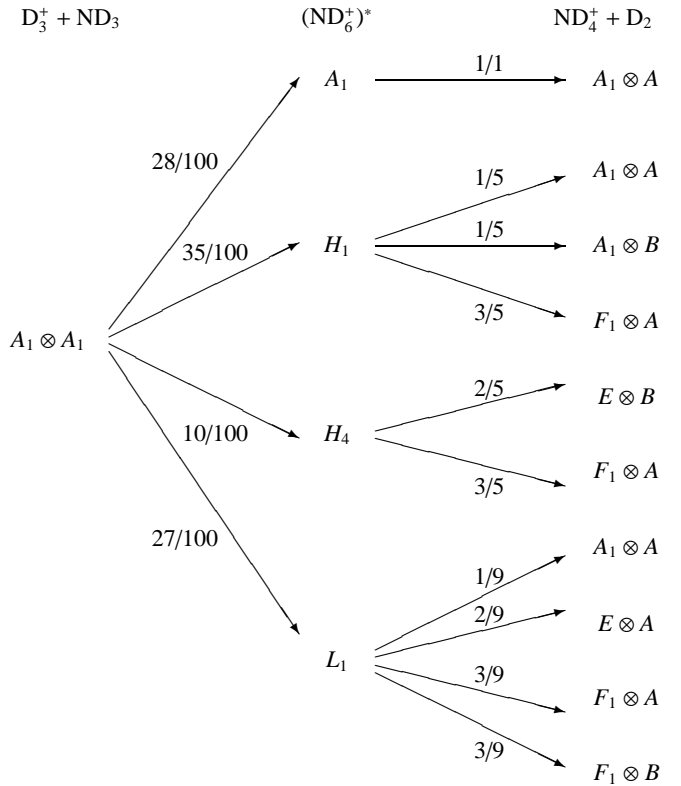

\end{document}